\RequirePackage{fix-cm}

\documentclass[twocolumn,final]{svjour3}

\def\mathbi#1{\textbf{\em #1}}
\smartqed  

\usepackage{graphicx}
\usepackage{amsmath}
\usepackage{setspace}

\begin{document}
\title{First-principles study of organically modified muscovite mica with ammonium (NH$_4^+$) or methylammonium (CH$_3$NH$_3^+$) ion}

\author{Chol-Jun Yu$^{1}$ \and Song-Hyok Choe$^1$ \and Yong-Man Jang$^2$ \and Gwang-Hyok Jang$^3$ \and Yong-Hyon Pae$^4$}

\institute{Chol-Jun Yu \\\email{ryongnam14@yahoo.com} \\ \\
$^1$ Department of Computational Materials Design, Faculty of Materials Science, Kim Il Sung University, Ryongnam-Dong, Taesong District, Pyongyang, Democratic People's Republic of Korea \\ \\
$^2$ Natural Science Centre, Kim Il Sung University, Pyongyang, Democratic People's Republic of Korea \\ \\
$^3$ Faculty of Geology, Kim Il Sung University, Pyongyang, Democratic People's Republic of Korea \\ \\
$^4$ Institute of Physics Engineering, Kim Chaek University of Technology, Pyongyang, Democratic People's Republic of Korea}

\date{Received: date / Accepted: date}

\maketitle

\begin{abstract}
Using density functional theory calculations, we have investigated the interlayer cation exchange phenomena in muscovite mica, which is motivated by a necessity to develop flexible high-insulating covering materials. The crystalline structures, chemical bonding properties, energetics, and electronic properties of muscovites before and after exchange of interlayer K$^+$ cation with ammonium (NH$_4^+$) or methylammonium (CH$_3$NH$_3^+$) ion were calculated. It was found that the unit cell volume changes are negligibly small upon exchange with NH$_4^+$ ion, while the unit cells are expanded with about 4 \% relative rate when replacing the interlayer K$^+$ cation with CH$_3$NH$_3^+$ ion. The energy band gap of pre-exchanged muscovite was calculated to be about 5 eV, which hardly changes upon interlayer cation exchange with either NH$_4^+$ or CH$_3$NH$_3^+$ ion, indicating the preservation of high insulating property of muscovite. The exchange energies were found to be about -100 kJ/mol for NH$_4^+$ and about -50 kJ/mol for CH$_3$NH$_3^+$ exchange, indicating that the exchange reactions are exothermic. A detailed analysis of atomic resolved density of states and electron density redistribution was provided.
\end{abstract}


\section{Introduction}\label{intro}
Muscovite mica, the most representative sheet silicate, has attracted extensive technological interest due to its extraordinary material properties such as high electrical insulation, thermal stability, and lubricity~\cite{Madhukar}. It is, for example, widely used as a pigment extender in various paints, an inert filler in rubber goods, a high temperature lubricant and flame-proof insulation of polyvinyl chloride (PVC) flexible insulating wires, and a reinforcing filler and extender in composite plastic systems~\cite{Gusev,Madhukar}. Its relative atomic smoothness~\cite{Ostendorf,Qi} and large band gap (7.85 eV)~\cite{Davidson,Kim} promise this material an excellent substrate for the growth of various materials including graphene~\cite{Krakow,Lamelas,Lippert11,Lippert13,Rudenko}. Recent reports demonstrated the fabrication of large area organic field-effect transistors based on a mica single crystal gate insulator (100 nm in thickness)~\cite{Gomez,He,McNeil}. In particular, organic modified muscovites, which are formed by exchange of potassium cation placed between dioctahedral aluminosilicate layers against organic ions like alkylammonium ion, have proved to be an admirable addition to significantly enhance mechanical stability in plastics and rubbers, and optical properties in thin films~\cite{Fujii,Heinz2,Heinz1,Osman2,Osman1}.

The structure of muscovite with chemical formula of KAl$_2$(Si$_3$Al)O$_{10}$(OH)$_2$ can be described as stacks of 2:1 layers, with each layer comprising AlO$_6$ octahedral sheet sandwiched between two SiO$_3$ tetrahedral sheets. Substitution of one Al$^{3+}$ cation for one of every four Si$^{4+}$ cations as shown in the chemical formula leads to the layers having a net negative charge, which is compensated by interlayer K$^+$ cations. Strong electropositive K$^+$ cations make natural muscovite hydrophilic and thus allow only the interaction with polar materials like water~\cite{Wang}. Organic modification of muscovite mica changes the polarity of the surface of the mica sheets from polar into nonpolar and, therefore, reverses the reactivity of mica sheets from hydrophilic into hydrophobic or even amphiphilic, resulting in the enhancement of the adhesion of mica to nonpolar materials due to the reduction of the interfacial free energy with the adjoining organic polymeric materials.

There have been several experimental~\cite{Fujii,Osman2,Osman1,Russell} and computational studies~\cite{Heinz2,Heinz1,Wang} for modification of muscovite with organic molecules. Heinz et al.~\cite{Heinz2,Heinz1} have developed an accurate extention of the consistent force field 91 (cff91) for mica and investigated mica structure and cation exchange capacity (CEC) with alkylammonium ions including 12 octadecyltrimethylammonium (C$_{18}$) ions or 12 dioctadecyldimethylammonium (2C$_{18}$) ions with periodic simulation cells by NVT molecular dynamics (MD). They have found the tilting of alkyl chains between layers, where the alkylammonium ions reside preferably above the cavities in the mica surface and the nitrogen atoms are 380 to 390 pm distant to the superficial Si-Al plane~\cite{Heinz2}. The MD simulations with different alkyl chains on mica surface indicated that long chains ($\geq$C$_{18}$) lead to a mixed phase of alkali ions and organic ions on the surface, medium chains ($\sim$C$_{12}$) give rise to phase separation, and very short chains prefer again homogeneously mixed surface~\cite{Heinz1}. The MD simulation work for water adsorbed muscovite surface has also been reported~\cite{Wang}. However, these MD simulations are not sufficient to give a valuable mechanism behind experimental observations at electronic scale.

In recent years, first-principles methods within the framework of density functional theory (DFT) have applied to the seires of phyllosilicate mica, yielding the reliable crystal structures, elastic and thermal properties~\cite{Hernandez,Militzer,Ortega,Ulian}. Surfaces and interfaces of muscovite mica were also studied by DFT calculations~\cite{Alvim,Vatti}, focusing on surface structure and images, and surface phase diagram in vacuum and in contact with water or an ionic liquid. Rosso et al.~\cite{Rosso} have performed {\it ab initio} calculations to investigate the structure and energetics of the Cs/K exchange into interlayer sites in muscovite, providing the fact that the Cs/K exchange rate and degree of irreversibility are likely to be dominated by diffusion kinetics due to the very low exchange energetics. Upon Na/K cation exchange, changes in the adhesion energy between muscovite surfaces~\cite{Sakuma13,Sakuma15} and in the elastic behavior between bulk muscovite models~\cite{Hernandez} were studied with DFT calculations. To the best our knowledge, however, first-principles study of organically modified muscovite has never been conducted so far; there are many remaining uncertainties with respect to the atomistic structure, energetics, and electronic properties upon interlayer cation exchange.

In this paper, we present a first-principles study of muscovite modified by interlayer K$^+$ cation exchange with ammonium (NH$_4^+$) or methylammonium (CH$_3$NH$_3^+$) ion with a wish to resolve these uncertainties. In Section~\ref{method}, we describe the computational method and modeling of bulk muscovite mica. In Section~\ref{result}, we provide the results from our DFT calculation with respect to the structure, energetics, electronic property and charge transfer upon K$^+$ exchange with above mentioned organic ions. Conclusion of the paper is given in Section~\ref{concl}.

\section{Method}\label{method}
\subsection{Structure and models}\label{struct_model}
The crystal of muscovite belongs to the monoclinic crystalline system with a space group of $C2/c$, where lattice parameters are like $a\neq b\neq c,~\alpha=\beta=90^\circ(\neq\gamma)$. There are four formula units ($Z=4$) in the conventional unit cell, which can be transformed to the primitive unit cell with lattice vectors $\mathbi{a}_p=0.5\mathbi{a}-0.5\mathbi{b},~\mathbi{b}_p = 0.5\mathbi{a} + 0.5\mathbi{b}$ and $\mathbi{c}_p=\mathbi{c}$, halving the formula units ($Z=2$). The structure of muscovite consists of aluminosilicate layers, so called 2:1 or T-O-T (Tetrahedral-Octahedral-Tetrahedral) layers, where silicons reside inside oxygen tetrahedra while aluminiums inside oxygen octahedra. The Al$\rightarrow$Si substitution for every four silicon atoms in the tetrahedral sites changes the charge property of T-O-T layer from neutral Al$_2$Si$_4$O$_{10}$(OH)$_2$ to negative charged Al$_2$(Si$_3$Al)O$_{10}$(OH)$_2$ that is balanced by interlayer K$^+$ cation. The tetrahedral sheets are linked each other to form siloxane rings. Among three sites in the octahedral sheet, two sites are occupied by Al atoms, and one site is vacant. A hydroxyl group is located at this vacant octahedral site, and K$^+$ cation is located at the siloxane cavity with 12-fold coordination.

\begin{figure}[!ht]
\begin{center}
\includegraphics[clip=true,scale=0.22]{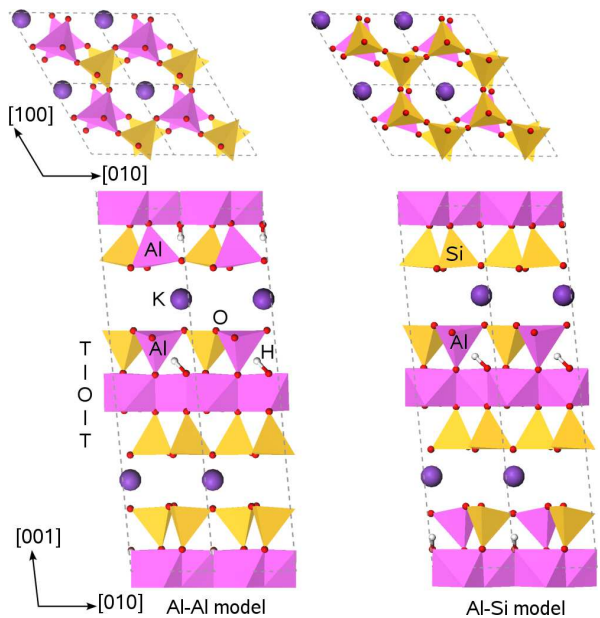}
\end{center}
\caption{\label{fig_model}(Color online) Polyhedral structure models for substitution of Si by Al in muscovite. Tetrahedral-Octahedral-Tetrahedral (T-O-T) layers are shown. In Al-Al model, two Si atoms at opposing sides of different T-O-T layers are substituted by Al atoms, while in Al-Si model the substitutions are occurred in the same sides.}
\end{figure}

Isomorphic substitution of Al$^{3+}$/Si$^{4+}$ leads to a local charge in the proximity of Al$^{3+}$ as well as structural disorder in the tetrahedral sheets. Since the short-range ordering in the distribution of Al/Si substitution is possible~\cite{Bailey} and Al$^{3+}$ cations arranged in the same sheet repel one another (so-called Loewenstein's rule prohibits Al$_\text{tet}$-O-Al$_\text{tet}$ linkages)~\cite{Militzer}, we consider two possible models for substitution in the primitive unit cell; the first model is constructed by replacing two Si atoms by Al at opposing sides of different T-O-T layers, while the second model has substitutions in the same sides. As shown in Fig.~\ref{fig_model}, these two models are denoted as Al-Al model and Al-Si model, respectively. Such substitutions transform the space group of muscovite into $C2$, and the occupation of a tetrahedral sheet into Si$_{0.5}$Al$_{0.5}$.

\begin{figure}[!ht]
\begin{center}
\includegraphics[clip=true,scale=0.3]{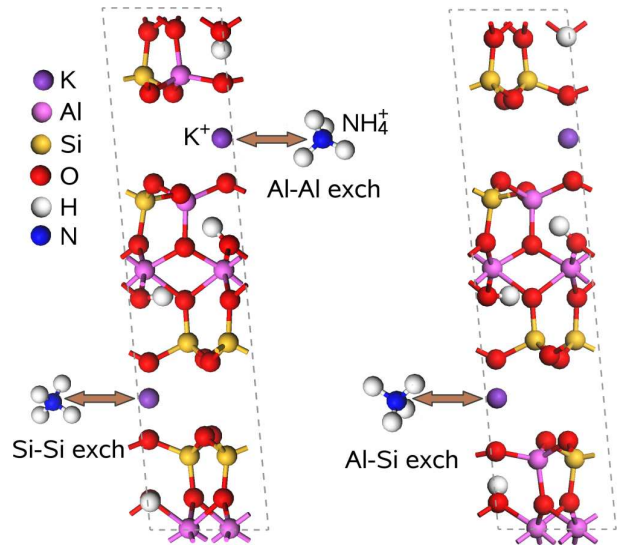}
\end{center}
\caption{\label{fig_exchmod}(Color online) Ball-and-stick models for cation exchange in muscovite. Al-Al, Si-Si, and Al-Si exchange models are possible according the interlayer K sites. One full interlayer sites are exchanged while another sites are remained.}
\end{figure}

Let us turn to the modeling for interlayer cation exchange. In this study, we consider the interlayer K$^+$ cation exchange by ammonium (NH$^+_4$) or methylammonium (CH$_3$NH$^+_3$) ion, which can be served as an initial step for long-length polymer modification of muscovite mica as well as fundamental physico-chemistry of organic modification of sheet silicates. Although the overall cation exchange at interlayer sites in sheet silicates can contain several steps such as delamination, diffusion and release of K, and sorption of other cations (see Refs.~\cite{Rosso,Shroll,Smith,Young}), we consider only the geometry-optimized solid phases before and after exchange, and isolated K$^+$, NH$_4^+$, and CH$_3$NH$_3^+$ cations. On the condition that we use the two models of muscovite with the primitive unit cell, three kinds of model are possible for cation exchange; (1) Al-Al exchange model where cation exchange occurs at the interlayer K site located at the centre of the upper and lower Al/Si di-trigonal six-member rings, and (2) Si-Si exchange model at the interlayer K site surrounded purely siloxane rings, and (3) Al-Si exchange model which is derived from the Al-Si model of muscovite. These three models are depicted in Fig.~\ref{fig_exchmod}.
Note that the cation exchange is occurred in one interlayer site while another K cation is remained in the primitive unit cell. For the case of methylammonium ion, the organic ion can be inserted into the interlayer space with parallel or perpendicular state to the basal plane of rings.

\subsection{DFT calculations}\label{dft_calc}
The full geometry optimizations were carried out by using \textsc{SIESTA} package~\cite{SIESTA}. We tested the validity of the approximations for the exchange-correlation interaction between valence electrons with the Perdew-Zunger type local density approximation (LDA) ~\cite{PZlda} and the Perdew-Burke-Ernzerhof type generalized gradient approximation (GGA)~\cite{PBE}. The Coulombic interactions between ionic core (nucleus plus core electrons) and valence electrons are described by Troullier-Martins~\cite{TMpseudo} type norm-conserving pseudopotentials, which were generated by ourselves within LDA and whose transferabilities were checked carefully. To expand the wave functions of valence electrons with localized numerical basis set, namely pseudo atomic orbitals, we used the double $\zeta$ plus polarization (DZP) orbitals. Mesh cutoff that defines the equivalent plane wave cutoff for the real space grid, and $k$-grid cutoff which determines the fineness of the reciprocal space grid used for Brillouin zone sampling were set to be 200 Ry and 5 \AA, respectively. All atoms were relaxed until the forces on each atom converge to 0.02 eV/\AA. These calculation parameters have already proved to give reliable results for the interaction between organic molecule and complicated silicate surfaces~\cite{yucj09,yucj16}.

It is well accepted that both LDA and GGA do not adequately describe long-range dispersive interaction, which is of much importance between the layers in sheet materials, as proved recently for 2$M_1$ muscovite by Ulian and Valdr\`{e}~\cite{Ulian} and for graphite by ourselves~\cite{yucj14}. To resolve this problem, the dispersive van der Waals (vdW) energy estimated by the semiempirical approach could be added to the DFT total energy of the system. In this work, we used the DFT-D2 method suggested by Grimme~\cite{DFT-D2}, in which the vdW energy is written as follows,
\begin{equation}
E_\text{vdW}=-s_6\sum_{j<i}\frac{C_6^{ij}}{r_{ij}^6}\frac{1}{1+e^{-d(r_{ij}/R_{ij}^{0}-1)}}
\end{equation}
Here, $r_{ij}$ is the distance between the $i$-th and the $j$-th atoms, $C_6^{ij}=\sqrt{C_6^iC_6^j}$ being $C_6^i$ the dispersion coefficient of the $i$-th atom, $R_{ij}^0=R_i^0+R_j^0$ being $R^0$ the vdW radius of atom, and $s_6$ and $d$ are scaling and damping parameters. The default values of 1.66 for $s_6$ and 20.0 for $d$ in SIESTA program, and the predetermined values for $C_6$ and $R^0$ of all atoms listed in Ref.~\cite{DFT-D2} were used in this work.~\footnote{We slightly modified SIESTA code to enable treatment of atom-pairs over ten as indicated in the SIESTA Academic License assigned to Yu.}

In order to estimate the lattice related properties like bulk modulus, we have calculated the total energies of unit cells with the volumes shifted systematically from the equilibrium volume $V_0$. The changes of volume $(V-V_0)/V_0$ were set to be $\pm$0.1. To do this, we have first performed the constant-volume optimization of unit cells at each fixed volume, and subsequently, fitted the calculated $E-V$ data to the Birch-Murnaghan equation of state~\cite{birch} for crystalline solids. We have made an analysis of Mulliken atomic charge populations.

In addition, we have calculated the electronic band structures and density of states with pseudopotential plane wave method implemented in Quantum ESPRESSO package~\cite{QE}, where ultrasoft pseudopotentials provided in the package were used for all atoms~\cite{uspp}. The kinetic energy cutoff for plane wave expansion was set to be 40 Ry and $k$-points for the first Brillouin zone integration to be (6$\times$6$\times$2). The vdW-DF4 method~\cite{vdwDF4} was used for vdW interactions.

\section{Result and discussion}\label{result}
\subsection{Structural properties}\label{res_struct}
The optimized lattice parameters of the pre-exchanged muscovite models are listed in Table~\ref{tab-lattice} (lattice parameters are given in the primitive unit cell). Although the optimizations were carried out without symmetry constraints, the structures approach the proper space group $C2$ within a maximal 0.004 \AA~search tolerance in length and a maximal 0.27$^\circ$ in angle. The obtained lattice parameters were in good agreement with the experimental data~\cite{Vaughan} for both Al-Al and Al-Si muscovite models. As general tendency in DFT calculations, GGA overestimates the lattice lengths within 3\% error to the experiment, while LDA provides values closer to the experiment (within 1\% error). It is worthy to note that whenever treating the vdW interactions the lattice lengths were slightly contracted compared to those obtained by LDA or GGA only.~\footnote{Note that localized basis sets like atomic orbitals might not properly consider the dispersive force compared to the expanded basis set like plane wave.} Although it was found that the Al-Si model is slightly lower in energetics than the Al-Al model with $\sim$ 0.1 eV/cell energy difference, we will use these two models for further investigation of interlayer cation exchange to consider the different local environments around interlayer cation.
\begin{table}[!ht]
\footnotesize
\caption{\label{tab-lattice}Optimized lattice parameters and unit cell volume $V_0$ of pre-exchanged muscovite models within LDA and GGA with and without DFT-D2 dispersion correction.}
\begin{tabular}{@{\hspace{5pt}}c@{\hspace{5pt}}c@{\hspace{5pt}}c@{\hspace{5pt}}c@{\hspace{5pt}}c@{\hspace{5pt}}c@{\hspace{5pt}}c@{\hspace{5pt}}c@{\hspace{5pt}}c}
\hline
 Model &$a$ (\AA)&$b$ (\AA)&$c$ (\AA)&$\alpha$ ($^\circ$)&$\beta$ ($^\circ$)&$\gamma$ ($^\circ$)&$V_0$ (\AA$^3$) \\
\hline
\multicolumn{8}{l}{(LDA)} \\
 Al-Al &5.180 &5.176 &19.966 &92.97 &93.04 &120.14 &460.37  \\
 Al-Si &5.180 &5.182 &20.010 &93.08 &93.09 &120.14 &461.82  \\
\multicolumn{8}{l}{(DFT-D2)} \\
 Al-Al &5.174 &5.170 &19.910 &92.97 &93.06 &120.15 &457.91  \\
 Al-Si &5.176 &5.177 &19.882 &93.07 &93.05 &120.14 &458.09  \\
\hline
\multicolumn{8}{l}{(GGA)} \\
 Al-Al &5.217 &5.217 &20.642 &93.14 &93.10 &120.10 &483.15  \\
 Al-Si &5.219 &5.222 &20.632 &93.47 &93.20 &120.11 &485.75  \\
\multicolumn{8}{l}{(DFT-D2)} \\
 Al-Al &5.208 &5.214 &20.589 &93.08 &93.11 &120.08 &480.97  \\
 Al-Si &5.214 &5.221 &20.710 &93.46 &93.20 &120.12 &484.36  \\
\hline
 Exp.$^a$ &5.165 &5.165 &20.071 &92.87 &92.87 &120.09 &460.96 \\
\hline
\end{tabular} \\
$^a$~Ref.~\cite{Vaughan}
\normalsize
\end{table}

In this work, the interlayer separation is defined and measured as $d_\text{int}=c(Z_{\text{O}_\text{h}}-Z_{\text{O}_\text{l}})\cos\beta$, where $c$ is length of lattice vector $\mathbi{c}$, $\beta$ is lattice angle, and $Z_{\text{O}_\text{h}}$ and $Z_{\text{O}_\text{l}}$ are fractional coordinates of lowest oxygen among three oxygens forming tetrahedra in the upper plane and highest oxygen in the lower plane, respectively. The interlayer separations depend on the local environment around interlayer cation. Among three types of local environments in the cases of pre-exchanged muscovite models, the largest separation was found to be 3.422 \AA~at the Al-Al environment where both upper and lower hexagonal rings contain Al atom, the smallest to be 2.805 \AA~at the Si-Si environment which do not contain Al atom in both rings, and the Al-Si environment has the between value of about 3.1 \AA, as listed in Table~\ref{tab-inter}. As mentioned above, Al/Si exchange in the tetragonal sheet leads to the layer having negative charge, and thus gives rise to the repulsive force between upper and lower layers. The tendency of interlayer separation change indicates that the attractive interaction between K$^+$ cation and negative charged layer can be overwhelmed by this repulsive interaction between the layers.
\begin{table}[!ht]
\caption{\label{tab-inter}Length of lattice vector $\mathbi{c}$, interlayer separation $d_\text{int}$, unit cell volume $V_0$ and bulk modulus $B$ of muscovite models before and after exchange by ammonium and methylammonium ions, optimized with LDA. Exch. means the exchanging particle, and ``aa'', ``ss'', and ``as'' denote Al-Al, Si-Si, and Al-Si local environments, respectively.}
\small
\begin{tabular}{r@{\hspace{8pt}}c@{\hspace{5pt}}c@{\hspace{5pt}}c@{\hspace{5pt}}c@{\hspace{5pt}}c}
\hline
  & $c$ & \multicolumn{2}{c}{$d_\text{int}$ (\AA)} & $V_0$ & $B$ \\
\cline{3-4}
 Model & (\AA) & K$^+$   & Exch. & (\AA$^3$) & (GPa) \\
\hline
 Al-Al &19.966 &2.805(ss) &  &460.37 &71.7 \\
       &       &3.422(aa) &  &       &     \\
 Al-Si &20.010 &3.119(as) &  &461.82 &57.1 \\
       &       &3.155(as) &  &       &     \\
\hline
       &       &          & NH$_4^+$    &       &         \\
 Al-Al exch &19.967 &2.821(ss) & 3.467(aa) &460.66 &63.3  \\
 Al-Si exch &19.782 &3.105(as) & 3.053(as) &460.33 &52.8  \\
 Si-Si exch &19.958 &3.440(aa) & 2.823(ss) &460.90 &68.8  \\
\hline
       &       &          & CH$_3$NH$_3^+$    &       &   \\
 Al-Al exch &20.965 &2.787(ss) & 4.407(aa) &482.42 & 62.1 \\
 Al-Si exch &21.133 &3.140(as) & 4.200(as) &487.66 & 49.0 \\
 Si-Si exch &20.927 &3.502(aa) & 4.058(ss) &485.22 & 59.3 \\
\hline
\end{tabular}
\normalsize
\end{table}

When exchanging one of two K$^+$ cations in the primitive unit cell with NH$_4^+$ ion, the lattice length $c$ and interlayer separation were found to be similar values to those of pre-exchanged muscovites, indicating the similar ionic radii between K$^+$ and NH$_4^+$ ions.

\begin{figure*}[!ht]
\begin{center}
\includegraphics[clip=true,scale=0.4]{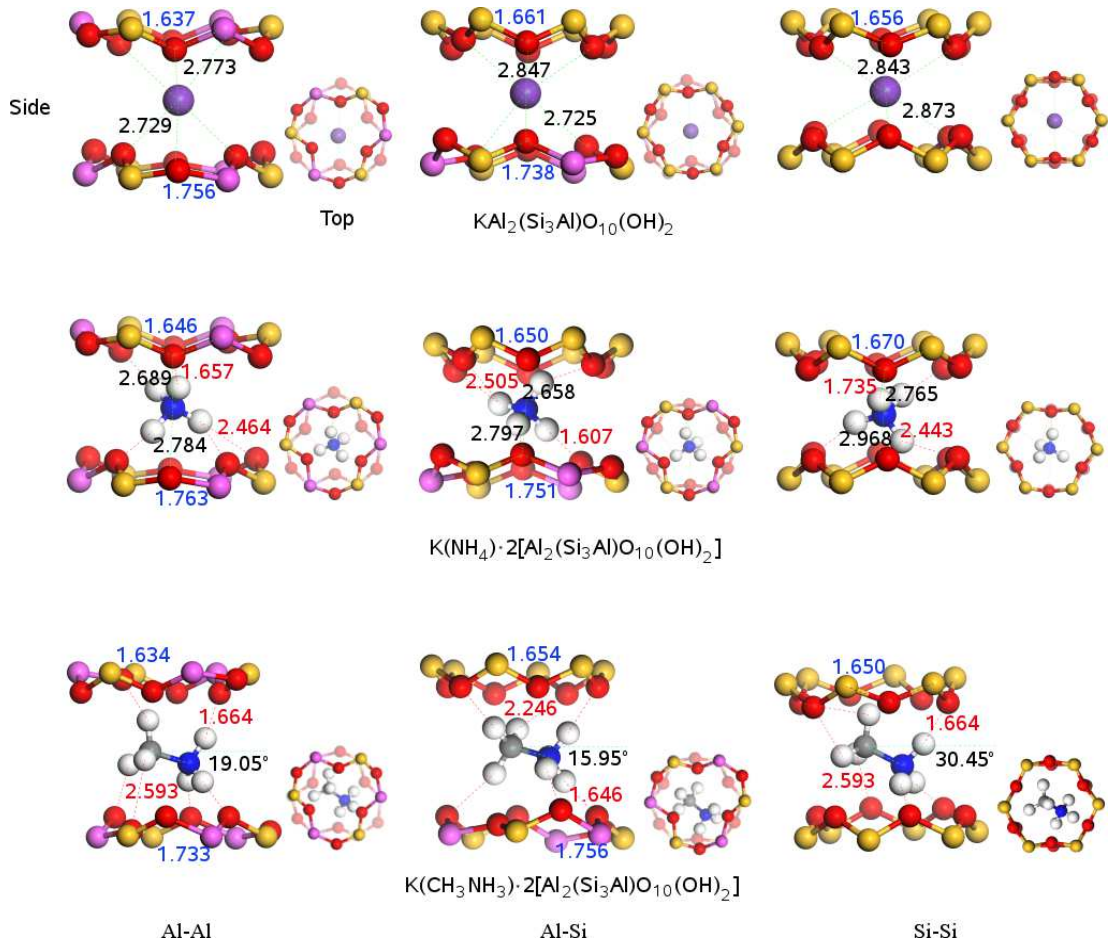}
\end{center}
\caption{\label{fig_bond}(Color online) Ball-and-stick illustration of local environments around interlayer cations of muscovite models before (top panel) and after cation exchange with NH$_4^+$ ion (middle panel) and CH$_3$NH$_3^+$ ion (bottom panel). Three kinds of local environments are possible; (1) Al-Al where upper and lower tetrahedral sheets contain Al atoms (left side), (2) Al-Si where only one sheet contains Al atom (centre column), and (3) Si-Si where both sheets do not contain Al atom (right side).}
\end{figure*}

In the case of CH$_3$NH$_3^+$/K$^+$ cation exchange, we can consider overall two different conformations of methylammonium ion with respect to the incline angle of its C-N axis (parallel and normal) to the basal plane of tetrahedral sheets. Moreover, it should be regarded also which layer side faces N atom between Al/Si replacing side and pure Si side in the case of Al-Si exchange model. It turns out that the parallel conformations are energetically lower than normal ones with about 0.4 eV/cell energy difference for three exchange models, and the conformations that N atom faces the Al-containing layer side and C atom towards the pure Si layer side is more favorable in energetics than its opposite conformations with about 0.6 eV/cell energy difference. On the basis of these observations, we present the most favorable conformations only in Table~\ref{tab-inter} and Fig.~\ref{fig_bond}. Upon CH$_3$NH$_3^+$/K$^+$ exchange, both lattice length and interlayer separation were enlarged as can be expected easily, resulting in the expansion of the unit cell volumes with 4.8\% and 5.6\% relative expansion rates over pre-exchanged muscovite models, as shown in Table~\ref{tab-inter}. For both NH$_4^+$/K$^+$ and CH$_3$NH$_3^+$/K$^+$ exchanges, the trends of interlayer separation increase according to the the local surroundings of interlayer cation are the same to the case of K$^+$ cation; Si-Si $<$ Al-Si $<$ Al-Al.

In Table~\ref{tab-inter}, we can also find the decrease of bulk modulus upon interlayer cation exchange. In the case of NH$_4^+$/K$^+$ exchange, the bulk modulus decreases from 71.7 GPa for pre-exchanged Al-Al muscovite model to 63.3 GPa for Al-Al exchange model or to 68.8 GPa for Si-Si exchange model, and from 57.1 GPa for pre-exchanged Al-Si muscovite model to 52.8 GPa for Al-Si exchange model. We can find further decrease of bulk modulus upon CH$_3$NH$_3^+$/K$^+$ exchange. The decrease of bulk modulus upon cation exchange indicates that muscovite becomes to be tender when modifying it with organic ions against interlayer K$^+$ cation.

The obtained Si-O bond lengths for all models before and after cation exchange varied from 1.634 \AA~to 1.670 \AA, which compares reasonably well with the known mean tetrahedral Si-O bond length of 1.647 \AA~for most silicate minerals~\cite{Gibbs,Guggenheim}, while the tetragonal Al-O bond lengths were estimated to be between 1.733 \AA~and 1.763 \AA, as shown in Fig.~\ref{fig_bond}. The distortions and tiltings of tetrahedra were observed, which can be thought due to the mismatch between Si-O and Al-O bond lengths. For the cases of pre-exchanged muscovite models in the top panel of Fig.~\ref{fig_bond}, the interlayer potassium cation is 12-fold coordinated with surrounding oxygen atoms with K-O bond lengths from 2.725 \AA~to 2.873 \AA~in good agreement with the experimental value of 2.848 \AA~\cite{Guggenheim}. Due to the attractive interaction between K$^+$ cation and negative charged layer created by Al/Si exchange in the tetrahedral sheets, the K$^+$ cation is more attracted to the Al-Si mixing layer than to the Si-Si layer, resulting in shortening of K-O bond lengths in the side of Al-Si layer (from 2.725 \AA~to 2.773 \AA) compared to the side of Si-Si layer (from 2.822 \AA~to 2.873 \AA).

Upon NH$_4^+$/K$^+$ exchange, resulting in the change of chemical formula to K(NH$_4$)$\cdot$2[Al$_2$(Si$_3$Al)O$_{10}$(OH)$_2$], the NH$_4^+$ cation is binded to the upper and lower layers with hydrogen bonds (bond lengths between 1.607 \AA~and 2.505 \AA) formed between H atoms of ammonium ion and O atoms of layers. The lengths between N atom and O atoms of layers were found to be between 2.659 \AA~and 2.968 \AA, which are comparable to those of K-O bond lengths in the pre-exchanged muscovite. In the cases of CH$_3$NH$_3^+$/K$^+$ exchange models, hydrogen bonds with bond lengths from 1.646 \AA~to 2.593 \AA~are also formed between methylammonium ion and tetrahedral sheets. The incline angles of C-N axis of the molecule to the basal plane were found to be 19.05$^\circ$ (Al-Al model), 15.95$^\circ$ (Al-Si model), and 30.45$^\circ$ (Si-Si model), although the organic ions were initially arranged in parallel to the basal plane.

\subsection{Exchange energetics}\label{res_energ}
We have estimated the interlayer cation exchange energetics, which can be calculated using the total energies for the optimized structures. For the NH$_4^+$/K$^+$ and CH$_3$NH$_3^+$/K$^+$ cation exchanges, the chemical reaction formula can be written as follows,
\begin{equation}
\label{eq:reaction1}
\text{X}(\text{K})+\text{NH}_4^+\Rightarrow \text{X}(\text{NH}_4)+\text{K}^+,
\end{equation}
\begin{equation}
\label{eq:reaction2}
\text{X}(\text{K})+\text{CH}_3\text{NH}_3^+\Rightarrow \text{X}(\text{CH}_3\text{NH}_3)+\text{K}^+,
\end{equation}
where X(K), X(NH)$_4$ and X(CH$_3$NH$_3$) are muscovites with K, NH$_4$ and CH$_3$NH$_3$ intercalates, respectively. The exchange energy can be calculated as a difference of Gibbs free energies of the corresponding systems. As discussed in Section~\ref{res_struct}, the volume changes ($\Delta V$) are quite small upon NH$_4^+$/K$^+$ and even CH$_3$NH$_3^+$/K$^+$ exchange, and, therefore, the $P\Delta V$ work term in the enthalpic part of the exchange energetics according to $\Delta H=\Delta E+P\Delta V$, where $\Delta E$ is the internal energy change, can be ignored. In fact, this term turned out to be about 3$\sim$4 orders of magnitude smaller than the internal energy difference. Moreover, we assume that the entropic contribution ($T\Delta S$) with the order of $k_{\text{B}}T$ at room temperature could be negligibly small compared to the internal energy difference with the order of few eV. Therefore, we can estimate the exchange energies as a balance of the DFT total energy change in muscovite and those of isolated ions. For the calculation of total energies of isolated ions, the simple cubic supercells with a lattice constant of 50 \AA~were adopted.

According to Table~\ref{tab_energ}, we can find that both NH$_4^+$/K$^+$ and CH$_3$NH$_3^+$/K$^+$ exchange reactions are exothermic, with the reaction energies of $\sim$ 100 kJ/mol for the former case and $\sim$ 50 kJ/mol for the latter case. The results suggest that the magnitude of exchange energy decreases with the increase of the size of interlayer organic ion, and thus for quite large or long molecules like 12 octadecyltrimethylammonium (C$_{18}$) polymer the reaction might become to be endothermic. It should be noted that some energy can be required to activate the chemical reactions in either endothermic or even exothermic way.
\begin{table}[!ht]
\caption{\label{tab_energ} Interlayer cation exchange energies according to the chemical reaction formula, equations \ref{eq:reaction1} and \ref{eq:reaction2}, and energy barriers for ion migrations as calculated by NEB method.}
\small
\begin{tabular}{rccc}
\hline
       & \multicolumn{2}{c}{Exchange energy} & Migration energy\\
\cline{2-3}
 Model & (eV) & (kJ/mol) & (eV) \\
\hline
       & \multicolumn{3}{c}{K$^+$} \\
 Al-Al & - & & 3.56 \\
 Si-Si & - & & 9.87 \\
 Al-Si & - & & 6.36 \\

\hline
       & \multicolumn{3}{c}{NH$_4^+$} \\
 Al-Al & -1.08 & -104.27 & 3.55 \\
 Si-Si & -1.06 & -102.34 & 8.09 \\
 Al-Si & -1.17 & -112.96 & 5.82 \\
\hline
       & \multicolumn{3}{c}{CH$_3$NH$_3^+$}\\
 Al-Al & -0.54 & -51.88 & 1.65 \\
 Si-Si & -0.58 & -56.15 & 4.61 \\
 Al-Si & -0.63 & -60.99 & 1.80 \\
\hline
\end{tabular}
\normalsize
\end{table}

\begin{figure}[!ht]
\begin{center}
\includegraphics[clip=true,scale=0.43]{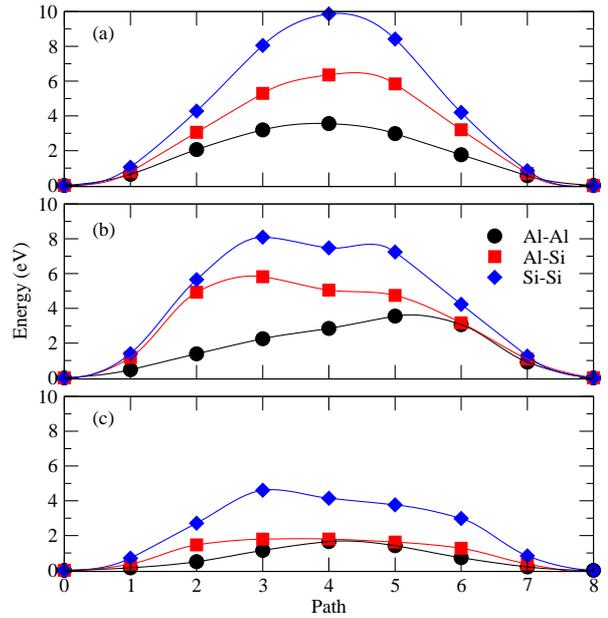}
\end{center}
\caption{\label{fig_neb}(Color online)Energy barriers for migration of (a) K$^+$, (b) NH$_4^+$, and (c) CH$_3$NH$_3^+$ ions in muscovite, calculated by using NEB method with vdW (DFT-D2) corrected LDA functional. Nine NEB images were used to construct a path.}
\end{figure}

In addition we have calculated the energy barriers for ion migration by using the nudged elastic band (NEB) method as implemented in SIESTA code in conjunction with Python script, Pastafarian. Two-times larger supercells were used to allow ions to migrate from one position to another identical position and nine NEB image configurations were used to discretize the path. The calculated migration energies listed in Table~\ref{tab_energ} are relatively high, indicating that in muscovite it is not easy to exchange potassium with other organic cations in external environment as other sheet silicates~\cite{Churakov}. As shown in Figure~\ref{fig_neb}, the local environment around interlayer cation affects the ion migration; Al-Al environment has the lowest energy barrier for ion migration, while Si-Si has the highest value.

\subsection{Electronic properties}\label{res_elect}
The electric insulating property of materials can be predicted by estimating a band gap through the calculation and analysis of the electronic energy band structure. A large band gap indicates a high insulating property. Therefore, we have calculated the electronic band structures of the systems within LDA and GGA, by employing both SIESTA~\cite{SIESTA} and Quantum ESPRESSO~\cite{QE} codes using the geometries optimized by SIESTA work. The calculated band gaps were varied from 3.96 eV to 5.02 eV according to the adopted method. For pre-exchanged muscovite, the values are severe lower than the experimental value of 7.85 eV~\cite{Davidson}, but it is not very astonishing when considering that LDA and GGA produce even half of the real band gap in most of insulators. Moreover, we do not aim to reproduce accurate band gaps of muscovite-related materials in this work though it would be possible using, for example, hybrid functionals, but to draw a valuable findings about cation exchange effect and its mechanism.

\begin{figure}[!ht]
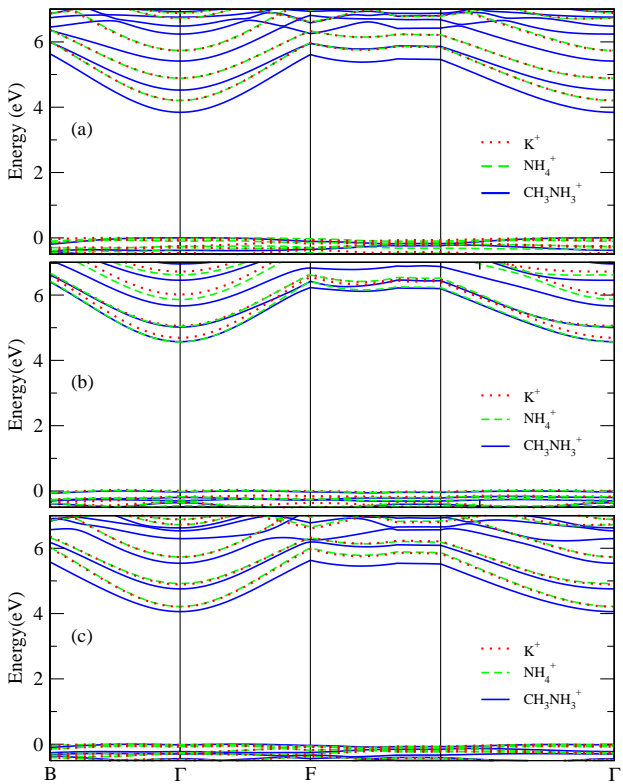

\begin{flushleft}
\includegraphics[clip=true,scale=0.43]{fig5a.eps} \\
\includegraphics[clip=true,scale=0.43]{fig5b.eps} \\
\includegraphics[clip=true,scale=0.43]{fig5c.eps}
\end{flushleft}
\caption{\label{fig_band}(Color online) Comparison of electronic band structures among pre-exchanged (red dotted line), NH$_4^+$/K$^+$ (green dashed line) and CH$_3$NH$_3^+$/K$^+$ (blue solid line) cation exchanged muscovites for (a) Al-Al, (b) Al-Si, and (c) Si-Si models. The valence band maximum is set to be zero.}
\end{figure}
In Fig.~\ref{fig_band}, we display a comparison of electronic band structures around band gap calculated with Quantum ESPRESSO code among pre-exchanged, NH$_4^+$/K$^+$ and CH$_3$NH$_3^+$/K$^+$ cation exchanged muscovites for Al-Al, Al-Si, and Si-Si models, respectively. It turns out that the band gaps are scarcely affected by interlayer cation exchange in spite of slight lowering of conduction band upon CH$_3$NH$_3^+$/K$^+$ exchange, and even they are identical in the case of Al-Si model. From these results, it can be concluded that the high insulating property of muscovite mica is kept after organically modification of interlayer cations, which is a positive aspect for making flexible high insulating wire by interlayer cation modification with organic ions.

\begin{figure}[!ht]
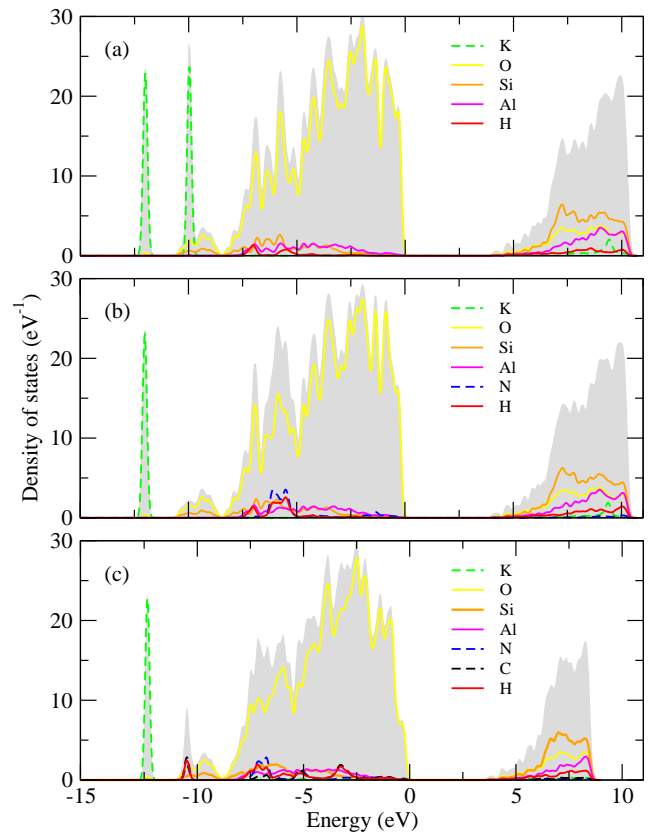

\begin{flushright}
\includegraphics[clip=true,scale=0.49]{fig6a.eps} \\
\includegraphics[clip=true,scale=0.49]{fig6b.eps} \\
\includegraphics[clip=true,scale=0.49]{fig6c.eps}
\end{flushright}
\caption{\label{fig_pdos}(Color online) Atomic resolved density of states of (a) pre-exchanged muscovite, (b) NH$_4^+$/K$^+$ and (c) CH$_3$NH$_3^+$/K$^+$ cation exchanged muscovite for Al-Al model. Total DOSs are plotted as filled grey colour.}
\end{figure}
In order to elucidate the role of atomic orbitals in electronic band structures, the atomic resolved density of states (DOS) were calculated. In Fig.~\ref{fig_pdos}, only the results for Al-Al model are presented, but the analysis for other models, Al-Si and Si-Si, is the same. In the valence bands below Fermi level, O (2p) states are predominant for all considered Al-Al models, that is, pre-exchanged, NH$_4^+$/K$^+$, and CH$_3$NH$_3^+$/K$^+$ cation exchanged muscovites, in one hand, and there are almost equivalent small contributions from other elements including Si, Al, C, N and H. On the other hand, we can see a hybridization between O (2p) states and mostly Si states and additionally Al states in the conduction bands above Fermi level. The peaks originated from the interlayer K atoms are placed far away from Fermi level, providing an implication that the interlayer K$^+$ cation has no influence on the magnitude of band gap and thus no change of insulating property of muscovite when occurring cation replacement by organic ions as mentioned in the discussion about energy band structure.

\subsection{Charge transfer}\label{res_charge}
\begin{figure}[!ht]
\begin{center}
\includegraphics[clip=true,scale=0.24]{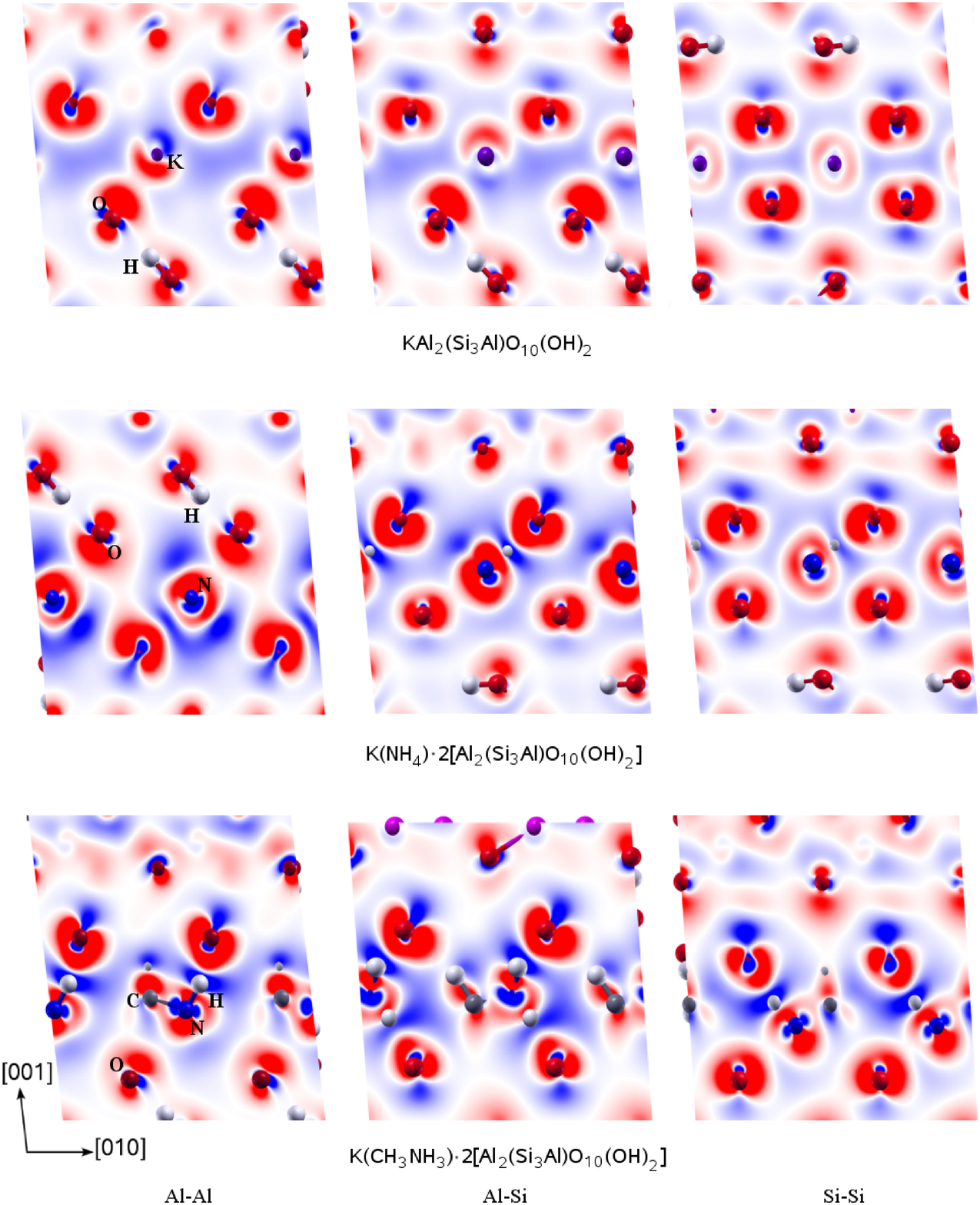}
\end{center}
\caption{\label{fig_dendiff}(Color online) Electron density redistribution around the interlayer cation in a plane perpendicular to the basal tetrahedral sheets in pre-exchanged (top panel), NH$_4^+$/K$^+$ (middle panel), and CH$_3$NH$_3^+$/K$^+$ (bottom panel) exchanged muscovites. Left, centre, right columns are for Al-Al, Al-Si, and Si-Si models, respectively. Electron density accumulation (depletion) is shown in red (blue).}
\end{figure}

To analyze the underlying mechanism, we discuss the bonding character based on the electron density redistribution and Mulliken charge population. The electron density redistribution ($\Delta\rho$) is defined as
\begin{equation}
\label{eq:deltarho}
\Delta\rho=\rho_{\text{X(int)}}-\rho_{\text{X}}-\rho_{\text{int}},
\end{equation}
where $\rho_{\text{X(int)}}$ and $\rho_{\text{X}}$ are the electron densities of the muscovite with and without interlayer cation (int), and $\rho_{\text{int}}$ is that of isolated interlayer cation, respectively. In the reference systems, the positions of the atoms correspond to the ones in the muscovite with interlayer cation. The electron density difference plots for the muscovites with interlayer cations K$^+$, NH$_4^+$, and CH$_3$NH$_3^+$ in Al-Al, Al-Si and Si-Si models are displayed in Fig.~\ref{fig_dendiff}. Only those in the local surroundings of the interlayer cations are shown.

In the case of pre-exchanged muscovite with the chemical formula KAl$_2$(Si$_3$Al)O$_{10}$(OH)$_2$, we find a strong accumulation of electron density around oxygens of both the upper and the lower tetrahedral sheets. Around the interlayer K$^+$ cations, accumulation in the side of lower sheet with hydrogen bonding to OH group and depletion in the side of upper sheet are shown in Al-Al model, accumulation in the side of pure Si tetrahedral sheet and week depletion in the side of Al and Si mixing sheet are observed in Al-Si model, and a week accumulation with a ring shape is shown in Si-Si model.

In NH$_4^+$/K$^+$ cation exchanged muscovite, strong charge depletion in the interstitial space and accumulation around O and N atoms are observed in Al-Al model. In Al-Si model, we can find a hydrogen bond formed between H of ammonium ion and O of tetrahedral sheet, indicating a strong binding between N and O through the hydrogen bond. In Si-Si model, similar observation is found. We confirmed that the hydrogen bonds are formed between hydrogen of methylammonium and oxygen of tetrahedral sheet in the case of CH$_3$NH$_3^+$/K$^+$ exchanged muscovite.

\begin{figure}[!ht]
\begin{center}
\includegraphics[clip=true,scale=0.35]{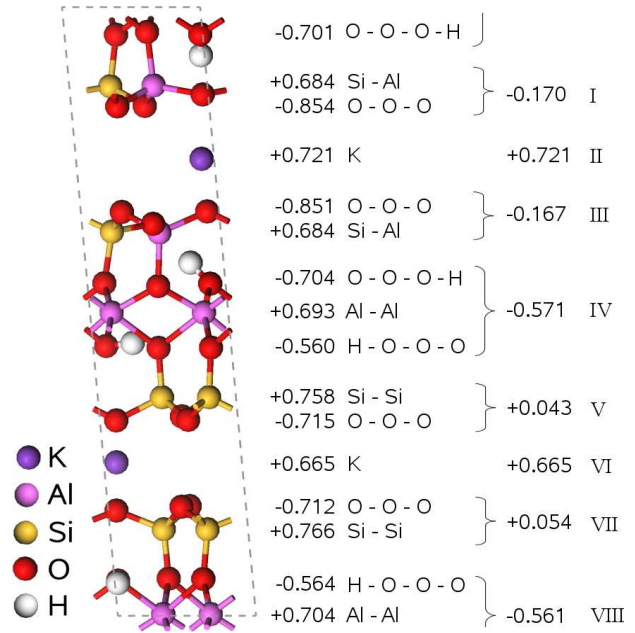}
\end{center}
\caption{\label{fig_mull}(Color online) Mulliken charge population in pre-exchanged Al-Al muscovite model. For the sake of simplicity, the 16 sheets in the primitive cell are rearranged in the 8 sheets by binding the neighboring ones.}
\end{figure}

We show the Mulliken charge population of the pre-exchanged Al-Al muscovite model in Fig.~\ref{fig_mull}, where totally 16 sheets in the primitive cell are rearranged in the 8 sheets by binding the neighboring ones for the sake of simplicity. It was found that the charge transferring in the atoms on the layer is not very much, and therefore the chemical bond in the layer is of covalent. The charges of the Al/Si exchanged sheets are negative, which is agreed with the prediction that Al/Si exchange leads to the layer having negative charge. The interlayer K atom is really close to ion, and thus has a strong electric binding with the negative charged layer. Concerning the charges of the sheets in the layer, the most of charge is accumulated in the middle part of the layer, while the charges on the surfaces of the layer are negligibly small. 
\begin{table}[!ht]
\caption{\label{tab_mull}Mulliken charge populations of pre-exchanged and NH$^+$/K$^+$ cation exchanged muscovites. Roman number of sheet is given in Fig.~\ref{fig_mull}.}
\small
\begin{tabular}{rrrrrr}
\hline
       & \multicolumn{2}{c}{Pre-exchanged} & \multicolumn{3}{c}{NH$^+$/K$^+$ exchanged}\\
\cline{2-3} \cline{4-6}
 Sheet & Al-Al & Al-Si & Al-Al & Al-Si & Si-Si \\
\hline
 I     & -0.170 & -0.270 & -0.056 & -0.300 & -0.180 \\
 II    &  0.721 &  0.689 &  0.494 &  0.692 &  0.722 \\
 III   & -0.167 &  0.150 & -0.085 &  0.168 & -0.169 \\
 IV    & -0.571 & -0.564 & -0.561 & -0.557 & -0.552 \\
 V     &  0.043 & -0.279 &  0.041 & -0.092 &  0.137 \\
 VI    &  0.655 &  0.694 &  0.662 &  0.475 &  0.494 \\
 VII   &  0.054 &  0.149 &  0.065 &  0.167 &  0.102 \\
 VIII  & -0.561 & -0.570 & -0.563 & -0.553 & -0.553 \\
\hline
\end{tabular}
\normalsize
\end{table}
In Table~\ref{tab_mull}, we provide the Mulliken charge populations rearranged according to the way depicted in Fig.~\ref{fig_mull} for the pre-exchanged Al-Al and Al-Si models, and NH$_4^+$/K$^+$ exchanged Al-Al, Al-Si, and Si-Si model. Similar arguments are found in other models. When substituting the interlayer K$^+$ cation by NH$_4^+$ ion, the amount of charge transferring on the interlayer site and its neighboring sheets is smaller than before exchange, while there is little change of charge population in other sheets upon the exchange, indicating that the interlayer cation exchange on one site hardly affects on the other neighboring interlayer site.

\section{Conclusion}\label{concl}
In conclusion, we have investigated the interlayer cation exchange effect in muscovite mica through DFT calculations. The interlayer K$^+$ cation is exchanged with organic ion, NH$_4^+$ or CH$_3$NH$_3^+$ ion, which can be thought to be useful for developing flexible high-insulating covering materials. Using the primitive unit cell that contains two formula units, the crystalline structures of pre-exchanged, NH$_4^+$/K$^+$, and CH$_3$NH$_3^+$/K$^+$ cation exchanged muscovites in each Al-Al, Al-Si, and Si-Si models have been optimized, providing a conclusion that the overall volume changes are negligibly small upon exchange with NH$_4^+$ but slight expansions are obtained in CH$_3$NH$_3^+$/K$^+$ exchange models, and organically modified muscovites become to be tender due to the decrease of bulk modulus. It is found that the interlayer distances are enlarged in both exchanged muscovites. From the electronic band structures, we have found that the band-gaps hardly change upon the interlayer cation exchange, which can be explained by placement of K states far below Fermi level through the analysis of atom resolved density of states. The calculated exchange energetics indicates that the interlayer cation exchange reactions are exothermic with about -100 kJ/mol for NH$_4^+$ exchange and about -50 kJ/mol for CH$_3$NH$_3^+$ exchange. The electron density redistribution is considered, providing an evidence of the charge transfer in the event of intercalation and the formation of hydrogen bonds between organic molecule and tetrahedral sheets. With all these findings, it can be concluded that it is feasible to exchange interlayer K$^+$ cation with organic ion, producing organically modified muscovites with a good flexibility and a high electric insulating property.

\section*{Acknowledgments}
The simulations have been carried out on the HP Blade System c7000 (HP BL460c) that is owned and managed by Faculty of Materials Science, Kim Il Sung University. This work was partially supported from the Committee of Education (grant number 02-2014), DPR Korea.

\section*{Conflict of interest}
The authors declare no competing financial interest.

\bibliographystyle{spmpsci}
\bibliography{Reference}

\end{document}